\documentclass[twocolumn,amsmath,floatfix,prb,aps]{revtex4}
\usepackage{color}
\usepackage{amsmath}
\usepackage{pifont}   % Ding symbols
\usepackage{graphicx} % Include figure files
\usepackage{dcolumn}  % Align table columns on decimal point
\usepackage{bm}       % bold math
\usepackage{amsfonts} % Some more fonts
\usepackage{amssymb}  % More symbols
\usepackage{multirow} % Table functions

\usepackage[english]{babel}
\usepackage[autostyle, english = american]{csquotes}
\MakeOuterQuote{"}

\usepackage{placeins}

\usepackage{braket} % as suggested by marmot and this example answer https://tex.stackexchange.com/questions/316836/the-bra-ket-notation-in-latex

\usepackage{bbm}

    \renewcommand{\v}[1]{\bm{\mathrm{#1}}}
    
    \newcommand{\tx}[1]{\text{#1}}

\begin{document}

\title{Ab-initio study of ultrafast charge dynamics in graphene}

\author{Q. Z. Li$^1$, P. Elliott$^1$, J. K. Dewhurst$^2$, S. Sharma$^1$ and S. Shallcross$^1$}
\affiliation{1 Max-Born-Institute for Non-linear Optics and and Short Pulse Spectroscopy, Max-Born Strasse 2A, 12489 Berlin, Germany}
\affiliation{2 Max-Planck-Institut fur Mikrostrukturphysik Weinberg 2, D-06120 Halle, Germany.}

\begin{abstract}
Monolayer graphene provides an ideal material to explore one of the fundamental light-field driven interference effects: Landau-Zener-St\"uckelberg interference. However, direct observation of the resulting interference patterns in momentum space has not proven possible, with Landau-Zener-St\"uckelberg interference observed only indirectly through optically induced residual currents. Here we show that the transient electron momentum density (EMD), an object that can easily be obtained in experiment, provides an excellent description of momentum resolved charge excitation. We employ state-of-the-art time-dependent density function theory calculations, demonstrating by direct comparison of EMD with conduction band occupancy, obtained from projecting the time propagated wavefunction onto the ground state, that the two quantities are in excellent agreement. For even the most intense laser pulses we find that the electron dynamics to be almost completely dominated by the $\pi$-band, with transitions to other bands strongly suppressed. Simple model based tight-binding approaches can thus be expected to provide an excellent description for the laser induced electron dynamics in graphene.
\end{abstract}

\maketitle

%\section{Introduction}

Intense laser light offers the possibility to control electrons in matter on femtosecond time scales. Triumphs of this burgeoning field include tuning the optically induced current in graphene via the carrier envelope phase of light\cite{Schiffrin2013,Heide2018,Higuchi2017}, attosecond control over magnetic order in thin films of magnetic overlayers\cite{dewhurst_laser-induced_2018,Siegrist2019}, and controlled valley excitation in the semi-conducting few layer dichalcogenides by circularly polarized light\cite{mak_control_2012,zeng_valley_2012} to name only a few examples. The two band Dirac cone found in graphene provides an ideal materials platform for studying one of the canonical light-field driven interference effects: Landau-Zener-St\"uckelberg (LZS) interference\cite{Shevchenko2010,Nori16_lzs}, which before its observation in graphene\cite{Higuchi2017} had only been observed in designed two state quantum systems\cite{DQD_current_lzs,GaAs_DQD_current_lzs,TiAu_qubit_current_lzs,Si_current_lzs,SiSiGe_QDQ_charge_transfer_lzs}. This effect occurs when an oscillating electromagnetic field drives intraband oscillation through the Bloch acceleration theorem $\v k \to \v k + \v A(t)/c$ and in the region of an avoided crossing interband transitions occur even when the band gap exceeds the dominant pulse frequency, so-called Landau-Zener transitions. Upon repeated passing of the avoided crossing multiple pathways exist to the conduction band with consequent constructive and destructive interference of electron states.
This offers rich possibilities for controlling electron dynamics by intense laser light, demonstrated by the recent observation of control over optical currents underpinned by LZS interference\cite{Higuchi2017}, a result anticipated theoretically in Ref.~\onlinecite{ishikawa_electronic_2013}.

The ubiquity of the avoided crossing band structure in 2d materials, found not only in the Dirac cone of graphene but also in the the semi-conducting monolayer dichalcogesides\cite{tmdc_bands}, phosphorene\cite{phosphorene_bands,Nematollahi2018}, silicene\cite{silicene_bands}, and stanene\cite{stanene_bands}, points towards the importance of LSZ interferometry in controlling electron dynamics in 2d materials. However, while interference physics can be easily probed theoretically through the conduction band population\cite{Nematollahi2018,Kelardeh2015,Kelardeh2016}, the experimental situation is more difficult, with to date only indirect observations of LSZ physics in materials reported. In this paper we show that the transient electron momentum density (EMD) difference, defined as

\begin{equation}\label{emd}
    \Delta\rho(\v p, t_f)=\rho(\v p, t_f)-\rho(\v p,t=0)
\end{equation}
with ${\bf p}$ momentum and $\rho(\v p,t)$ the electron momentum density\cite{dugdale} before ($t=0$) and after ($t_f$) the pump laser pulse, offers a tool for directly probing LZS interference effects. The EMD may be measured experimentally via 
tomographic reconstruction using Compton profiles\cite{STBK95,DH96,SSWK96,KSHH03,HBHA07} and, in particular, for layered materials\cite{HN17,emd3d_Kurp1996}. Combining these techniques with ultrafast X-ray sources will allow the transient EMD to be experimentally measured. This suggests a way in which the LZS physics may be directly observed in 2d materials, opening the way to correlate indirect LZS physics such as induced currents with the fundamental momentum space interference patterns. 

For graphene, we demonstrate that the EMD facilitates both the real time observation of the formation of LZS interference patterns, as well as the elucidation of subtle features in the relation between pump pulse and interference in momentum space.

In contrast to previous works that have employed simple single particle tight-binding Hamiltonians to study the LZS effect\cite{Nematollahi2018,Higuchi2017,Kelardeh2014,Kelardeh2015,Kelardeh2016,fillion-gourdeau_time-domain_2016,lefebvre_carrier-envelope_2018,gagnon_pulse_2018}, we will here deploy the time dependent version of density functional theory (TD-DFT).
To establish the accuracy of the EMD as a record of LZS interference we compare it with the excited electron distribution, $N_{\tx{ex}}$, defined within TD-DFT as\cite{elliott_ultrafast_2016}:

\begin{equation}\label{nex}
N_{\tx{ex}}(\v k, t) = \sum_i^{occ}\sum_j^{unocc}\left| \langle \psi_{i\v k}(t)|\psi_{j\v k}(t=0) \rangle\right|^2
\end{equation}
where $\psi_{j\v k}(t)$ is the time-dependent Kohn-Sham orbital at time $t$, and $\psi_{i\v k}(t=0)$ is the ground state orbital.
In all cases we find that the pattern of excitation in momentum space generated by transient EMD and N$_{\rm ex}$ is nearly identical in the first BZ.

Finally, we consider the role of the non-$\pi$-band states in the electron dynamics in graphene. Remarkably, despite electron excitation through the whole energy range of the $\pi$-band (up to 10~eV above the Fermi energy, an energy range encompassing the $\sigma^\ast$ bands as well as several high $l$ character bands), it turns out that there occur almost no transitions to states outside the $\pi$-band manifold. We attribute this to the near vanishing of the corresponding dipole matrix elements. Our calculations thus suggest that even for very significant laser excitation tight-binding based models will provide a good description of the electron dynamics.

%\section{Method}

\begin{figure}[t]
\includegraphics[width=0.5\textwidth]{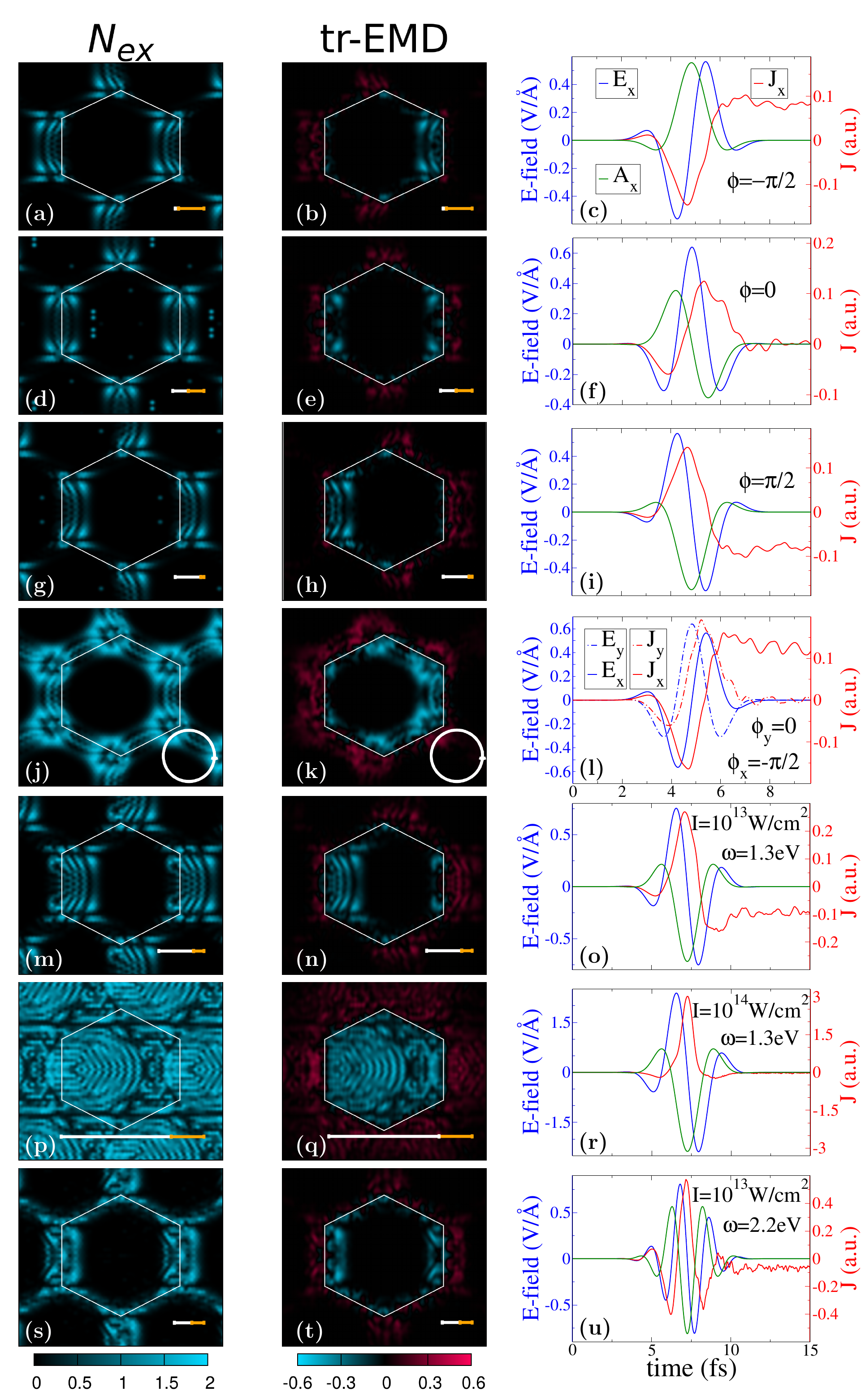}
\caption{{Conduction band occupation as a function of $\v k$-vector as determined directly by projection of the time-dependent state onto the ground-state Kohn-Sham states (first column), see Eq.~\eqref{nex}, and, second column, the transient electron momentum density (tr-EMD) difference, see Eq.~\eqref{emd}. Evidently, both quantities in a consistent way capture the momentum space intensity fringes generated by Landau-Zener-St\"uckelberg interference.
The third column displays the electric field ($\v E$-field) of the pump laser pulse (blue lines), the $\v A$-field scaled such that it can be plotted on the same axis (green lines), and the induced current density (red lines). Pulses in (a)-(l) have a full width half maximum (FWHM) of 1.935~fs, a central frequency of 1.4~eV, and peak intensity of $5.43\times10^{12}$~W/cm$^2$, and carrier envelope phase as indicated in the panels. The remaining three rows have FWHM 2.758~fs, CEP of $\pi/2$, and central frequencies and intensities as indicated in the panels.
In the first and second columns, the white hexagons represent the boundary of the 1st BZ while the lines in the right bottom corner represent the effective $\v k$-space trajectory given by the Bloch acceleration theorem. 
}}
\label{fig:cep}
\end{figure}

According to Runge-Gross theorem\cite{RG1984}, which extends the Hohenberg-Kohn theorem into the time domain, with common initial states there will be a one to one correspondence between the time-dependent external potentials and densities\cite{Carsten_book,TDDFTbook}.
Based on this theorem, a system of non-interacting particles can be chosen such that the density of this non-interacting system is equal to that of the interacting system for all times, with the wave function of this non-interacting system represented by a Slater determinant of single-particle orbitals. These time-dependent Kohn-Sham (KS) orbitals are governed by the Schr\"odinger equation (for the spin degenerate case):

\begin{eqnarray}
i\partial_t \psi_j({\bf r},t)=
\Bigg[
    \frac{1}{2}\left(-i{\nabla} +\frac{1}{c}{\bf A}_{\rm ext}(t)\right)^2 + v_{s}({\bf r},t)    \Bigg]
\psi_j({\bf r},t). \nonumber \\
\label{tdks}
\end{eqnarray}
In the above equation ${\bf A}_{\rm ext}(t)$ is the vector potential representing the applied laser field, the effective potential $v_s({\bf r},t)$ is given by $v_{s}({\bf r},t) = v_{\rm ext}({\bf r},t)+v_{\rm H}({\bf r},t)+v_{\rm xc}({\bf r},t)$, where $v_{\rm ext}({\bf r},t)$ is the external potential, $v_{\rm H}({\bf r},t)$ the Hartree potential, and $v_{\rm xc}({\bf r},t)$ is the exchange-correlation (xc) potential. For the latter we have used the adiabatic local density approximation. From the Fourier transform of the Kohn-Sham states, $\psi_{i\v k}({\bf r})$, the electron momentum density can be constructed as $\rho(\v p)=\sum_{i\v k}|\psi_{i\v k}(\v p)|^{2}$. 
This EMD constructed from KS states has been found to provide excellent agreement with that obtained from Compton scattering\cite{dugdale}.

All calculations employ the state-of-the art all-electron full potential linearized augmented plane wave (LAPW) method\cite{singh}, as implemented in the ELK code\cite{elk}. We have used a $30 \times 30$ {\bf k}-point set; for further details of the implementation of TD-DFT within the LAPW basis we refer the reader to Refs.~\onlinecite{kriegerjctc} and \onlinecite{dewhurst15}.

%\section{Results}

\begin{figure}[t]
    \includegraphics[width=0.45\textwidth]{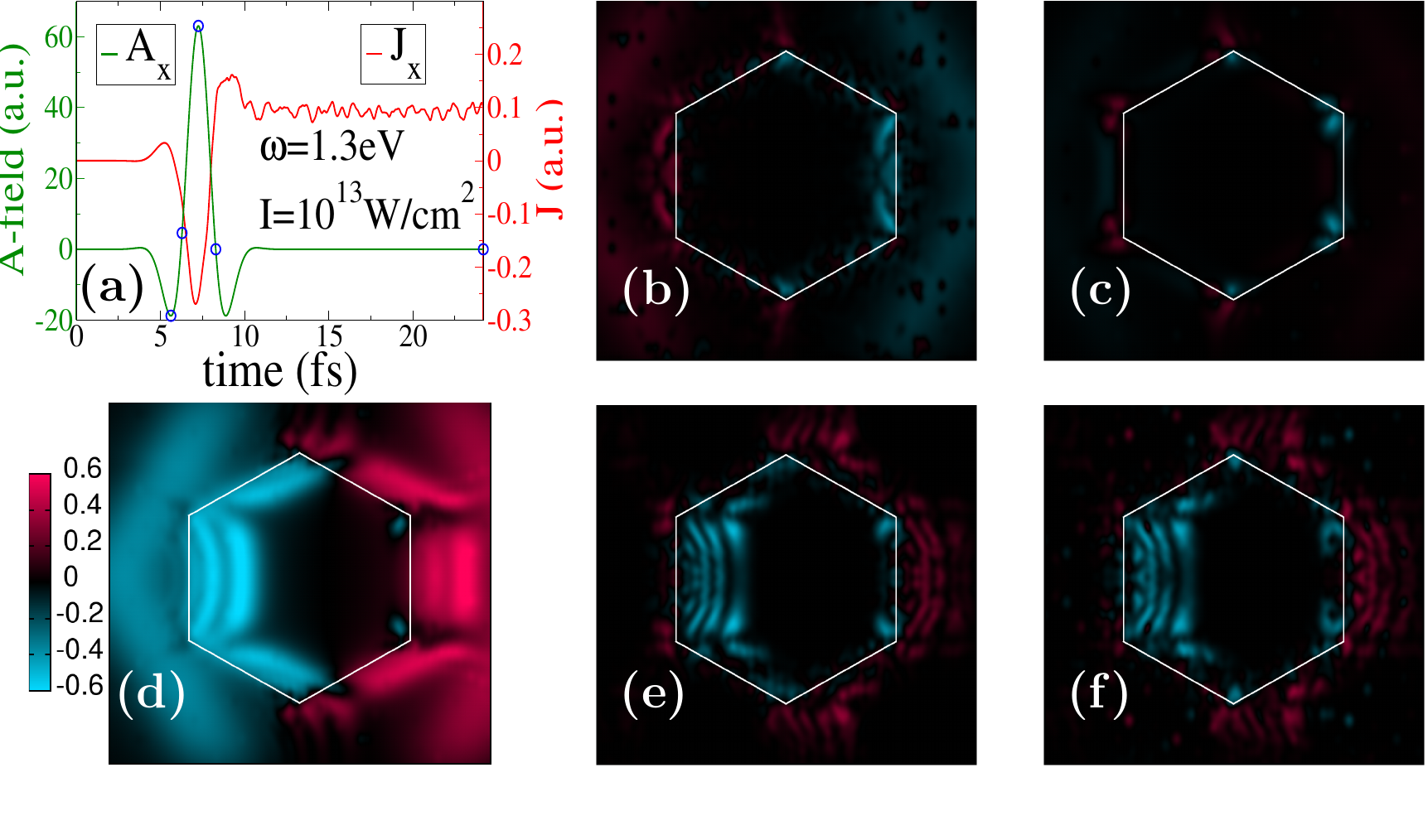}
    %\vspace{-0.5cm}
    \caption{\footnotesize{Landau-Zener-St\"uckelberg interference in the first Brillouin zone (BZ) reflected by 2D transient electron momentum density (tr-EMD) at various time steps during and after the pulse. A pulse of central frequency 1.3~eV, intensity $1.0\times10^{13}$~W/cm$^{2}$, full width half maximum 2.758~fs, and carrier envelope phase $\pi/2$ is employed, with the $\v A$-field exhibited and laser induced current exhibited in panel (a). The points on the $\v A$-field curve indicate the times at which the tr-EMD is evaluated, shown in panels (b-f). In these panels the full evolution of the Landau-Zener-St\"uckelberg (LZS) interference can be seen, including both early time $k_x < 0$ (left hand side of the vertical BZ boundary line) conduction band excitation, intense excitation at the pulse peak, panel (d), before the development of the LZS interference fringes on the falling shoulder of the pulse, panel (e), and the full time $k_x > 0$ LZS interference.}}
    \label{fig:int}
\end{figure}

{\it LZS interference probed by 2D tr-EMD}: the patterns of excited charge in momentum space that most directly characterise Landau-Zener-St\"uckelberg interference are generally presented by plotting the conduction band occupation over the first Brillouin zone. However this information, while easy to obtain theoretically, is difficult to obtain experimentally. We thus look at an alternative quantity, the change in electron momentum density due to the laser pulse.

In Fig.~\ref{fig:cep} are displayed the $N_{\rm ex}$, EMD, and induced currents for a diverse set of laser pulses exhibiting variation of several pulse parameters: carrier envelope phase (the angular difference between the $\v E$-field and pulse envelope maxima), polarization, intensity, frequency, and full width half maximum (FWHM). The magnitude of the electric field is of the order of 5~V/nm, placing these pulses in the strong non-perturbative regime for graphene. As can be seen, in all cases $N_{\rm ex}$ and EMD convey consistent information concerning the excited charge, establishing the latter as a reliable probe of momentum space excitation.

Before exploring the LZS interference physics of graphene revealed in Fig.~\ref{fig:cep} we first provide a theoretical basis to this observed coincidence in the pattern of momentum space excitation between $N_{\rm ex}$ and EMD.
The Kohn-Sham electron momentum density is defined as
\begin{equation}\label{emd_def}
\rho(\v p,t) = \sum_{j\v k} f_{j\v k} |\psi_{j\v k}(\v p,t)|^2
\end{equation}
where
\begin{equation}\label{phi_ft}
\psi_{j\v k}(\v p,t) = \int d\v p \ e^{i\v p\cdot\v r}\psi_{j\v k}(\v r,t)
\end{equation}
is the Fourier transform of the KS wavefunction $\psi_{j\v k}$ and $f_{j\v k}$ the occupation. Upon expansion of the Bloch functions in plane waves of the reciprocal lattice vectors, $\v G$,
\begin{equation}\label{phi_r}
%\psi_{j\v k}(\v r) = e^{i\v k\cdot\v r}u_{j\v k}(\v r)
\psi_{j\v k}(\v r,t) = \sum_{\v G} c^{\v k}_{j\v G}(t) e^{i(\v k + \v G)\cdot\v r}
\end{equation}
and insertion into Eqs.~(\ref{emd_def}) and (\ref{phi_ft}), we find that the EMD can be expressed as
\begin{equation}\label{emd_delta}
\rho(\v p,t) = \sum_{j\v k} f_{j\v k} \sum_{\v G} |c^{\v k}_{j\v G}(t)|^2 \delta(\v p - \v k - \v G).
\end{equation}
The EMD will therefore only change with respect to the ground-state EMD (see Eq. (\ref{emd})) at points $\v p = \v k +\v G$ where the coefficients $c^{\v k}_{j\v G}(t)$ change. In particular, in almost all systems, this will include the $\v G = 0$ point, i.e. the $\v k$ point itself within the first Brillouin zone. As the coefficient, $c^{\v k}_{j\v G=0}(t_f)$, will change (w.r.t the GS value) at points in k-space where $N_{\tx{ex}}(\v k, t_f)$ is non-zero, $\Delta\rho(\v k, t_f)$ must also then be non-zero. Hence, any interference pattern seen in $N_{\tx{ex}}(\v k, t_f)$ will also be seen in $\Delta\rho(\v k, t_f)$. For 2 electron systems, it is known that the EMD produced from the KS wavefunction can differ significantly from the exact EMD\cite{RHGM09,EM11}, however in periodic systems, it was shown that the KS-EMD gives excellent agreement with Compton Scattering profiles\cite{dugdale}.

For a carrier envelope phase (CEP) of $\phi = \pm \pi/2$ the maximum $\v E$-field intensity, and hence the interband transition at the avoided crossing, occurs at turning point of the path in momentum space executed due to the $\v A$-field. As a result, the LZ transitions generate excited conduction band charge at either the positive ($\phi=+\pi/2$) or negative ($\phi=-\pi/2$) $k_x$ sides of the Dirac point. This can be seen in rows (a-c) and (h-j) of Fig.~\ref{fig:cep}. Note that positive and negative $k_x$, measured from the Dirac point, corresponds to the left and right hand sides of the vertical BZ boundary as seen in Fig. 1. In contrast, for $\phi=0$ the maximum $\v E$-field intensity occurs at $\v A = \v 0$ resulting in a symmetric excitation about the Dirac point, see row (c-e). In the past such asymmetric LZS interference has been indirectly accessed by means the net current that results from the asymmetric momentum space occupation for non-zero CEP, and to date this represents the only observation in experiment of LSZ in a material\cite{Higuchi2017}. This coherent current (current per unit cell) induced by the laser pulse is displayed in the third column of Figs.~\ref{fig:cep} and \ref{fig:int}, and corresponds well with that seen in experiment. The experimentally accessible EMD, however, provides a wealth of additional information, as we now describe.

By comparing rows (g-i) and (m-o) we observe almost identical residual coherent current, and yet a very different momentum space LZS excitation as revealed by the EMD. 
%For the first pulse the FWHM is 1.935~fs, central frequency 1.4~eV and peak pulse intensity $5.43\times10^{12}$~W/cm$^2$, while for the second pulse the FWHM is 2.758~fs, central frequency 1.3~eV, and a peak pulse intensity of $10^{13}$~W/cm$^2$. Both pulses have have CEP of $+\pi/2$ however the EMD displays striking differences.
In particular, in row (m-o) we observe a sub-dominant $k_x < 0$ charge excitation absent in row (g-i) and reflecting multiple passes of the avoided crossing due to the side peaks of the former pulse, see panel (i). The presence of the main and side peaks in the pulse structure allows for multiple pass $\v k$-space trajectories which, due to the pulse envelope, consist of a series of passes of the avoided crossing from trajectories of different length in momentum space. This yields both asymmetric occupation and more complex interference patterns. Further enhancement of these side peaks, see row (s-u) of Fig. 1 for which the CEP is again $\phi=+\pi/2$, results in well developed interference fringes both for positive and negative $k_x$, quite different to the right hand side only momentum space occupation seen for the lower frequency $\phi=+\pi/2$ pulse shown in panels (g-i). The EMD thus represents a much more sensitive probe of the LZS effect, able to unveil subtitles of the interference physics lost in the residual current.

A striking example of this richness of information provided by EMD versus the residual current can be found in rows (p-r). Here it can be seen that widespread excitation occurs throughout the BZ driven by an intense pulse the $\v A$-field of which drives trajectories right across the BZ (indicated by the lines in the $N_{\rm ex}$/EMD panels). A very complex and asymmetric LZS interference pattern results from this intense excitation, however the widespread occupation of momentum space drives an overall cancellation of current carrying states and a vanishingly small residual current.

Experimentally, the short time coherent current ultimately generates heating and a diffusive residual current. This can therefore not provide a real time probe of the development of LSZ physics. Transient EMD, on the other hand, potentially provides a real time probe of the ultrashort time evolution of LZS interference patterns in momentum space. In Fig.~\ref{fig:int} we show the transient EMD evaluated before, during, and after a pump pulse inducing both a coherent current and LZS interference. One can observe an early time excitation due to pulse side peaks, panels (b) and (c), followed by a dramatic excitation in momentum space at the maximum of the main peak, panel (d). Only after this peak has passed do the final interference fringes develop, panels (e) and (f).

\begin{figure}
    \centering
    \includegraphics[width=0.45\textwidth]{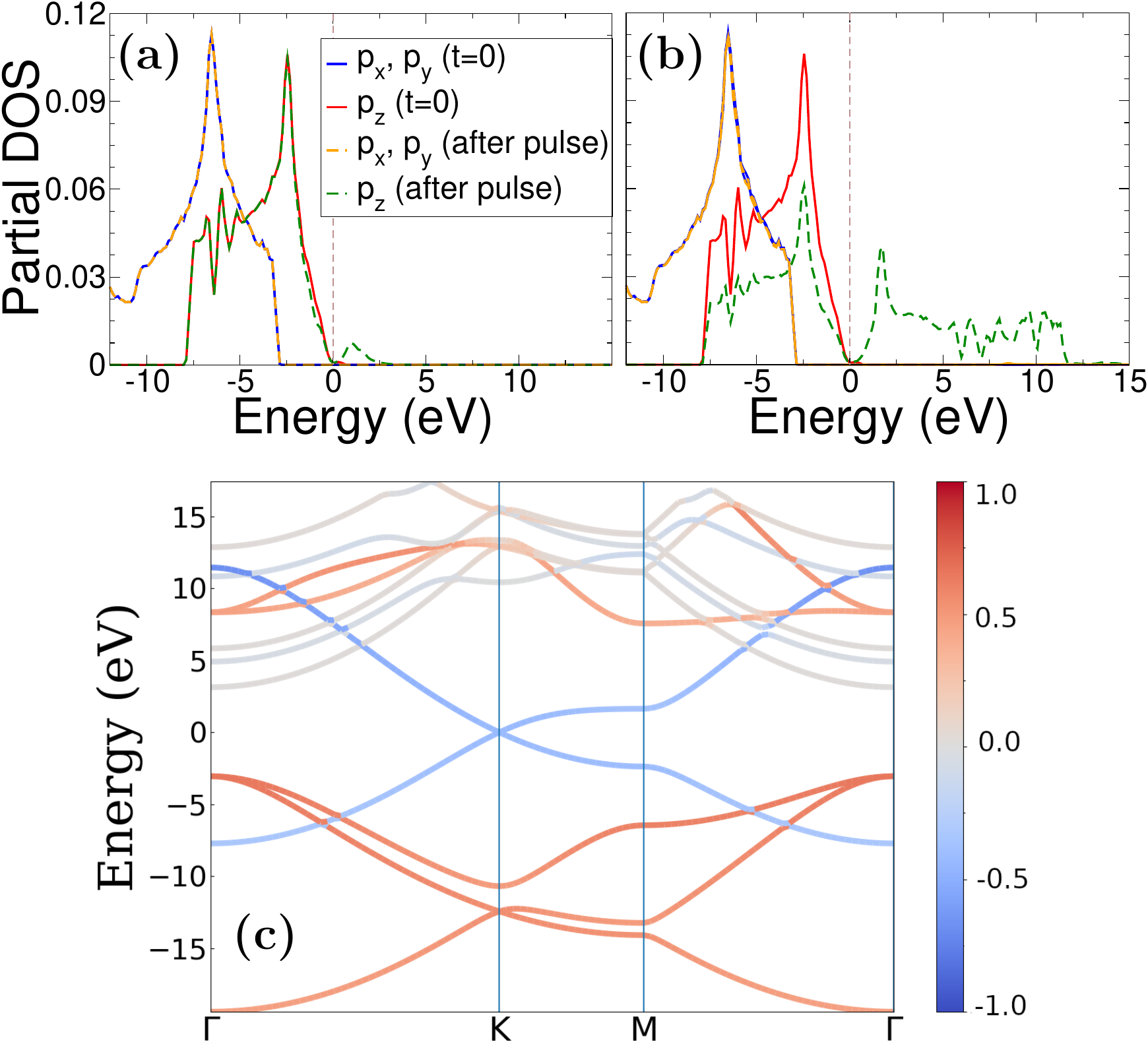}
    \caption{\footnotesize{Time dependent partial density of states (PDOS) projected onto the $l=1$ spherical harmonics. Here the PDOS (in states/atom/eV) is shown both at $t=0$ before the pulse, and at the end of the silulation after the pulse has been applied. The pump pulse for panels (a) and (b) is polarized in the $x$-direction, with intensities $10^{12}$~W/cm$^{2}$ and $10^{14}$~W/cm$^{2}$ respectively. As can be seen, even for almost complete excitation of the $\pi$-band in which charge is excited from the $\pi$-band minima up to the $\pi^\ast$-band maxima, there is no excitation into states of $p_x$ or $p_y$ character.
    (c) Band structure of graphene showing the $\pi$ and $\sigma$ band character. Negative and positive numbers indicate dominance by $\pi$- and $\sigma$-character respectively.}}
    \label{fig:pdos}
\end{figure}

{\it Dominance of the $\pi$-manifold in electron dynamics}:
the results for the momentum resolved conduction band occupation shown in the previous sections, correspond very closely to results obtained on the basis of model $\pi$-band only tight-binding Hamiltonians. This raises the question of whether this is due simply to the relatively low energies of the excited charge (in Fig~\ref{fig:cep} and Fig.~\ref{fig:int} the excited charge resides predominantly at the K point and the K-M-K line) or whether, for a more general reason, the $\pi$-band will always dominate ultrafast laser induced electron dynamics in graphene.
To explore this in Fig.~\ref{fig:pdos} we display the partial density of states calculated before and after the laser pulse. As can be seen, see Fig.~\ref{fig:pdos}a, for the pulse of intensity $10^{12}$~W/cm$^{2}$, the partial DOS after the pulse shows conduction band occupation only up to 2.5~eV. At these energies, see Fig.~\ref{fig:pdos}c, only the $\pi$-band is available for excited charge. Remarkably, when we consider a very strong pulse of intensity $10^{14}$~W/cm$^{2}$ the excited electrons are again only of $p_z$ character,  Fig.~\ref{fig:pdos}b, despite the fact that the laser pulse is sufficiently strong to excite charge from the minima of the $\pi$-manifold up to the maxima of the $\pi^\ast$-manifold. As may be noted from the band structure, Fig.~\ref{fig:pdos}c, within this energy range exist many other bands that would, in principle, be expected to be involved in the electron dynamics at such high energies.
Examination of the relevant dipole matrix elements reveals that transitions from $\pi$ to $\sigma^\ast$ and $\pi^\ast$ to $\sigma$ are negligible for laser pulses with in-plane polarization. Thus even in the highly non-perturbative regime transitions from the ground state to the $\sigma^\ast$ manifold will be strongly suppressed. It might be argued that the partial DOS, a projection within (touching) muffin tins, does not account for excitation to delocalized bands of high $l$ character. Comparison of the interstitial density of states before and after the pulse shows that there is indeed an increase in interstitial charge at around 9~eV, possibly indicating transitions from the $\pi^\ast$ manifold to delocalized bands (note the intersections between $\pi^\ast$ and high $l$ character bands on the M-$\Gamma$ line), however this is a rather small effect. It would thus appear that the model $\pi$-band only tight-binding Hamiltonians provide an excellent description of the electron dynamics even for very intense laser pulses.

%\section{Conclusions and discussion}
To summarize we have investigated \emph{ab-initio} the laser induced electron dynamics in monolayer graphene. This system provides a canonical example of a material for which Landau-Zener-St\"uckelberg interferometry can be explored, and we have shown that direct visualisation of the interference fringes in momentum space is possible via the transient electron momentum density (EMD), establishing transient EMD as an excellent experimental tool for exploring LZS interference in 2d materials.
Examination of the excited state partial density of states reveals that the
$\pi$-band manifold decisively dominates ultrafast laser induced dynamics in graphene, justifying the deployment of the popular H\"uckel tight-binding model. Whether this remains true for the complex few layer graphene systems, for which such an approach is the only one that can reasonably be envisioned, remains an open question.
\section{Acknowledgements}
QZL would like to thank DFG for funding through TRR227 (project A04). SS would like to thank DFG for funding through SH498/4-1 and PE acknowledges funding from DFG Eigene Stelle project 2059421. The authors acknowledge the North-German Supercomputing Alliance (HLRN) for providing HPC resources that have contributed to the research results reported in this paper.

\bibliographystyle{unsrt}
\bibliography{gra_paper}

\end{document}